\documentclass{osa-article}

\journal{oe}

\newcommand{\figref}[2]{Fig.~\ref{#1}~{(\bf #2)}}
\newcommand{\figrefs}[1]{Fig.~\ref{#1}}
\usepackage{graphicx}
\usepackage{amsmath}
\usepackage{color}
\usepackage{gensymb}
\usepackage{subfig}
\usepackage{upgreek}

\articletype{Research Article}

\begin{document}

\title{Single-shot Dispersion Sampling for Optical Pulse Reconstruction}

\author{A.~Korobenko,\authormark{1} 
P.~Rosenberger,\authormark{2,3}
J.~Sch\"otz,\authormark{2,3}
A.Yu.~Naumov,\authormark{1}
D.M.~Villeneuve,\authormark{1}
M.F.~Kling,\authormark{2,3}
A.~Staudte,\authormark{1}
P.B.~Corkum,\authormark{1}
and B.~Bergues\authormark{1,2,3,*}}

\address{\authormark{1}Joint Attosecond Science Laboratory, National Research Council of Canada and University of Ottawa, Ottawa, Ontario K1A0R6, Canada\\
\authormark{2}Physics Department, Ludwig-Maximilians-Universit\"at Munich, Am Coulombwall 1, 85748 Garching, Germany\\
\authormark{3}Max Planck Institute of Quantum Optics, Hans-Kopfermann-Stra\ss{}e 1, 85748 Garching, Germany}

\email{\authormark{*}boris.bergues@mpq.mpg.de} 



\begin{abstract}
We present a novel approach to single-shot characterization of the spectral phase of broadband laser pulses. Our method is inexpensive, insensitive to alignment and combines the simplicity and robustness of the dispersion scan technique, that does not require spatio-temporal pulse overlap, with the advantages of single-shot pulse characterization methods such as single-shot frequency-resolved optical gating at a real-time reconstruction rate of several Hz. 
\end{abstract}

\section{Introduction}

The continuous progress in ultra-fast laser science and technology has created a demand for simple and reliable pulse characterization methods. In frequency space, the temporal evolution of a laser pulse is completely determined by its spectral amplitude and phase. While the spectral amplitude is readily obtained using a spectrometer, measuring the spectral phase is more challenging. Techniques with varying complexity have been developed over the past years to determine the spectral phase of short laser pulses~\cite{Kane1993,Iaconis1999, Oshea2001, Theberge2006,  Moulet2010, Miranda2012, Ziv2020, Hammond2018, Park2018, Saito2018, Korobenko2020}. Popular methods such as FROG~\cite{Kane1993} and SPIDER~\cite{Iaconis1999} use a copy of the pulse to provide a reference. The spectral phase of the pulse is recovered by analysing the spectrum of a wave-mixing process as function of the delay between the pulse and the reference. Recently, a new method, called dispersion scan was proposed~\cite{Miranda2012}. Here, instead of using a separate pulse, the spectral components of different wavelengths are referenced to each other, and the variable time delay is substituted with a variable dispersion element - usually a glass wedge moved in and out of the beam path. The technique comes with the great advantage of a significantly simplified measurement because it does not require a spatio-temporal overlap of two pulses as other techniques.

Another important feature that distinguishes different pulse characterization techniques is their ability to reconstruct the temporal evolution of a single laser pulse. Although the original dispersion scan was designed as a multi-shot technique, several single-shot modifications have been implemented~\cite{Fabris2015,Louisy2017,Salgado-Remacha2020,Sytcevich2020}. Here, we demonstrate a novel method in which several reflections of the same pulse will experience each a different amount of dispersion. We implemented the differential evolution (DE) algorithm~\cite{Das2011, Escoto2017}, shown before to be well suited for dispersion scan trace reconstruction, on a consumer-grade graphics card. We demonstrate that a few reflections are enough to obtain a reliable reconstruction, which allows us to achieve a real-time pulse reconstruction rate of a few Hertz with an efficient parallel algorithm. Compared to competing techniques, our method is inexpensive, fast, robust and insensitive to the beam profile inhomogeneity.

\section{Single-shot dispersion scan}

\begin{figure}[ht!]
\centering\includegraphics{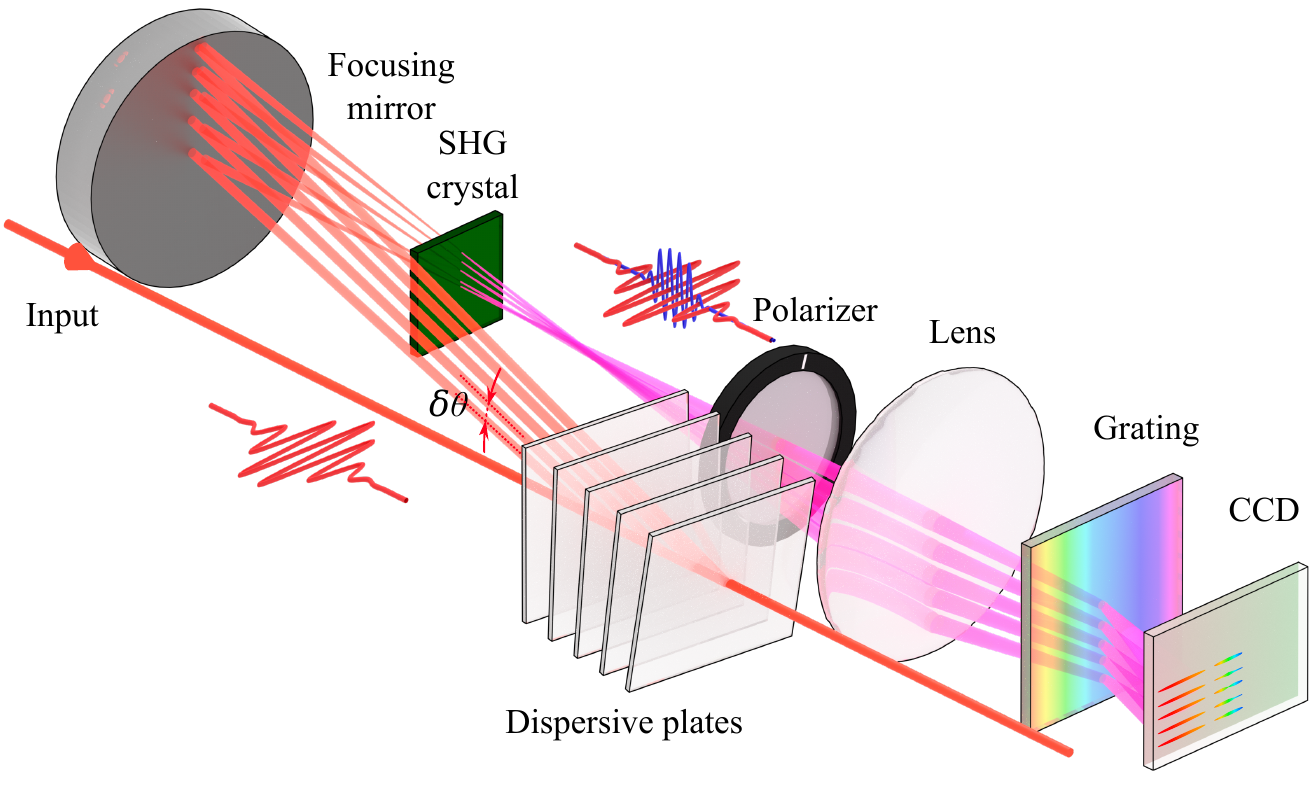}
\caption{Experimental setup for the dispersion sampling method. SHG: second harmonic generation, CCD: charge-coupled device of a camera.\label{fig:setup}}
\end{figure}

The idea is illustrated in \figrefs{fig:setup}. The input beam is polarized horizontally, and passes through a series of $M$ dispersive material plates. While the first plate is aligned normal to the beam, every consecutive one is rotated by a small angle $\delta\theta$ relatively to the previous one. This results in a fan of beams reflected in a backward direction. The beam reflected from the front surface of the $m$-th plate, $0\leq m<M$, propagates at an angle of $\theta_m=2m\delta\theta$ with respect to the horizontal. Its spectral amplitude and phase of $A_m(\omega)$ and $\phi_m(\omega)$ at the radial frequency $\omega$ are described as: 

\begin{gather}
    A_m(\omega) = A_0(\omega)\left\vert r_m(\omega)\prod_{m^\prime=0}^{m-1}t_{m^\prime}(\omega)\right\vert =  A_0(\omega)\rho_m(\omega)\\
    \phi_m(\omega) = \phi_0(\omega)+\arg\left(r_m(\omega)\prod_{m^\prime=0}^{m-1}t_{m^\prime}(\omega)\right)+\sum_{m^\prime=0}^{m-1}\frac{n_{m^\prime}(\omega)\omega}{c}d_{m^\prime}=\phi_0(\omega)+\varphi_m(\omega).
\end{gather}

In the equations above, $A_0(\omega)$ and $\phi_0(\omega)$ are the input pulse spectral amplitude and phase, respectively, and $r_m(\omega)$ is the front surface amplitude reflection coefficient, $t_m(\omega)$ is the front and the back surface amplitude transmission coefficient, $n_m(\omega)$ is the index of refraction, and $d_m(\omega)$ is the thickness of the $m$-th plate. The symbols $\rho_m$ and $\varphi_m$ were introduced for convenience. They describe the combined effect of the dispersive plates on the amplitude ($\rho_m$) and phase ($\varphi_m$) of the $m$-th beam.

The beams are then focused on a second harmonic generation (SHG) crystal to produce vertically polarized second harmonic radiation. The beams are dispersed in the horizontal plane with a diffraction grating, and focused with a lens onto the sensor of a CCD camera that images $M$ spectral traces, each containing the fundamental and the corresponding second harmonic pulse. A broadband polarizer with its axis oriented close to the vertical direction is used to attenuate the fundamental to the level of the second harmonic. A reconstruction algorithm is applied to recover the input pulse spectral phase $\phi_0(\omega)$ from the known $\rho_m(\omega)$ and $\varphi_m(\omega)$ and the measured spectral intensities $I_m^\mathrm{fund}$ and $I_m^\mathrm{SHG}$ of fundamental and second harmonic laser pulses, respectively.

It was shown in \cite{Escoto2017} that the differential evolution (DE) algorithm can be used for quick and efficient reconstruction of dispersion scan traces. Although the authors applied it to a conventional dispersion scan where the spectral intensity of the fundamental was independent of the amount of inserted glass, i.e. $\rho_m(\omega)=\mathrm{const}(m, \omega)$, the algorithm can be extended to the more general case of $\rho_m(\omega)\neq\mathrm{const}(m, \omega)$. Note that neither the thicknesses nor the material of the dispersive material plates have to be the same for the method to work. 

The ability to combine the plates of different thicknesses and/or materials extends the versatility of a conventional dispersion scan. It is especially useful for designing a universal device facilitating the reconstruction of laser pulses over a wide spectral range. Indeed, insertion of the same thickness of a dielectric can cause too much of a chirp at one wavelength and too little at the other, depending on the material dispersion, making the retrieval algorithm inefficient in both cases.  
Combining plates of different thicknesses and materials can provide effective dispersion for pulses in a broad range of bandwidths and wavelengths. While investigating the optimal combinations is of high practical interest, in the present proof-of-principle study we concentrated on the simplest case of a set of identical plates. In this case, $\rho_m(\omega)=\left\vert r(\omega)(1-r^2(\omega))^m\right\vert$, and $\varphi_m(\omega)=\pi+\frac{n(\omega)\omega m}{c}d$, where $r(\omega)=\frac{1-n(\omega)}{1+n(\omega)}$. We neglect the wavelength dependence of reflectivity $r(\omega)=r$, and normalize the measured second harmonic spectra according to $I_m^\mathrm{SHG, norm} = I_m^\mathrm{SHG}/\rho_m^2$. The problem then reduces to the reconstruction of a trace of a conventional dispersion scan with a constant step size\cite{Wnuk2016} $2d$, and the existing DE reconstruction algorithm can be applied directly without modifications.

We implemented the (DE) algorithm, outlined in \cite{Escoto2017} on a graphics processing unit (GPU) Nvidia 1070 TI, using CUDA technology. The approach involves finding the fundamental laser spectral phase $\phi(\omega)$, which, together with its measured amplitude $A(\omega)$ minimizes the error between the measured $I^\mathrm{meas}_{m}(\omega)$ and reconstructed $I^\mathrm{rec}_{m}(\omega)= \left\vert\mathfrak{F}\left\{\mathfrak{F}^{-1}\left\{A(\omega)e^{i(\phi(\omega) + \varphi_m(\omega))}\right\}^2\right\}\right\vert^2$ dispersion scan traces. The frequency $\omega$ was discretized with $N$ points: $\omega_n=\omega^\mathrm{min} + n(\omega^\mathrm{max}-\omega^\mathrm{min})/(N-1),~0\leq n<N$. The best phase was chosen out of a population of $D$ possible phases $\phi_{dn},~0\leq d<D$, after letting the initial population of random phases evolve for several thousands of generations. Following ref.~\cite{Escoto2017}, the evolution laws were chosen as follows. For each member of the population $\phi_{dn}$, we randomly selected three members of the remaining population $\phi_{an},~\phi_{bn},~\phi_{cn}$, and constructed a mutant member from them, according to $\phi^\mathrm{mut}_{dn} = \phi_{an}+\alpha_d(\phi_{bn}-\phi_{cn})$, where $\alpha_d$ is a random number between 0 and 1. The offspring $\phi^\mathrm{off}_{dn}$ was then generated from the initial $\phi_{dn}$ and a mutant $\phi^\mathrm{mut}_{dn}$ members, according to:
\begin{equation*}
\phi^\mathrm{off}_{dn} = \begin{cases}
\phi_{dn}, &\beta_{dn}\leq0.5\\
\phi^\mathrm{mut}_{dn}, &\beta_{dn}>0.5
\end{cases},
\end{equation*}
where $\beta_{dn}$ is a random number between 0 and 1. Out of the set of $2D$ phases $\left\{\phi_{dn}\right\}\cup\left\{\phi^\mathrm{off}_{dn}\right\}$, we selected the $D$ phases yielding the smallest error between $I^\mathrm{rec}_{mn}$ and $I^\mathrm{meas}_{mn}$. These phases were used as the next generation.

To improve the convergence, we gradually increased the wavelength grid size $N$ from $16$ to $128$ in each run. Parallel execution allowed us to increase the population size $D$ to 128 and with single-precision floating point arithmetic, the calculation rate reached tens of thousands of generations per second (GPS), yielding a few reconstructions per second, depending on the number of spectra $M$.

\begin{figure}[ht!]
\centering\includegraphics{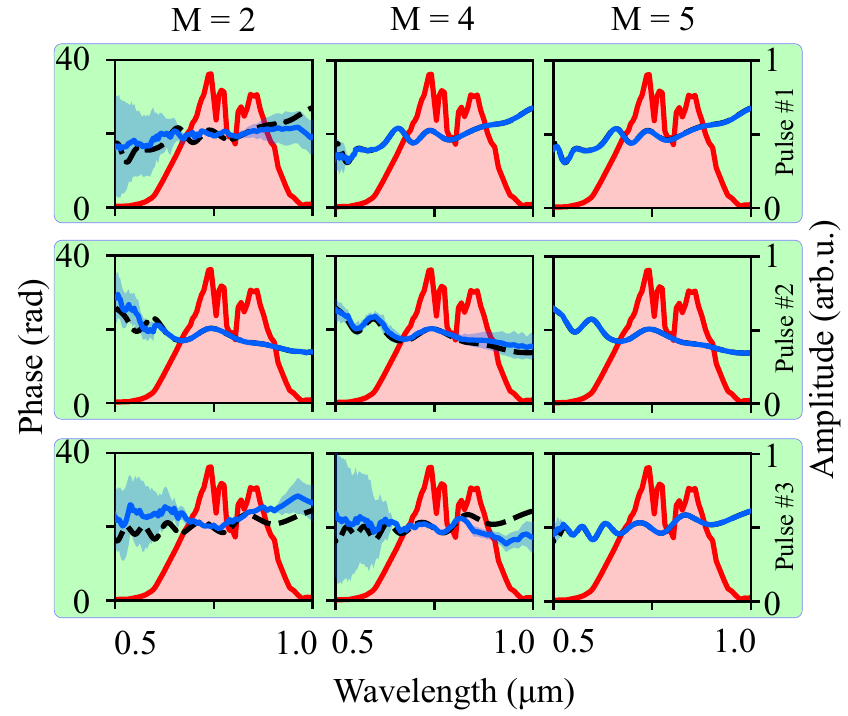}
\caption{Conversions of the differential evolution method for different number of plates $M$ (columns) and for a set of three randomly generated pulses (rows). Red-shaded shape represents the pulse amplitude spectrum, taken from a spectrometer measurement of a real pulse, the dashed black is a randomly generated phase, the solid blue line and light blue shading represent an average and range of phases, reconstructed in 10 consecutive runs, with 20000 generations per run, respectively.\label{fig:theor}}
\end{figure}

The choice of the optimal value of $M$ is a trade-off between the speed of the retrieval and the simplicity of the setup on the one hand, and the accuracy of the reconstruction on the other hand. In order to benchmark our method against different values of $M$, we performed simulations using an artificial pulse generated from the measured spectral amplitude of a Ti:Sapphire (Ti:Sa) output that was spectrally broadened in a hollow-core-fiber to a bandwidth of a 6~fs-long Fourier-Transform-limited pulse (cf. solid red line in \figrefs{fig:theor}) and a randomly generated smooth spectral phase. We first assigned a phase to 20 frequencies, uniformly covering the whole frequency range, randomly picking it from the interval [0, $2\pi$], and then used cubic splines to interpolate it to the intermediate points. Three examples of such random phase are shown in the three rows of \figrefs{fig:theor} with the black dashed lines, after subtraction of an insignificant linear phase. For each pulse we calculated the discrete dispersion scan spectrogram for several different values of $M$ and used our DE algorithm implementation to reconstruct the spectral phase. Repeating the reconstruction process 10 times, allowed us to estimate the accuracy of the retrieval procedure and extract an error bar for the spectral phase. We found $M=5$, corresponding to the last column in \figrefs{fig:theor}, to be an optimal value, allowing for an accurate reconstruction. At a reconstruction rate of 12000~GPS, we found the threshold goodness $G$~\cite{Miranda2012} set at $10^{-3}$ to be reached in less then 100~ms. The goodness reflects the difference between the true and reconstructed traces, and is defined as:

$$G=\sqrt{\frac{1}{NM}\sum_{m=0}^{M-1}\sum_{n=0}^{N-1}\left(I^\mathrm{meas}_{mn}-R(\omega_n)I^\mathrm{rec}_{mn}\right)^2},$$

where $R(\omega)$ is the spectral sensitivity of the measurement~\cite{Miranda2012}.

\begin{figure}[ht!]
\centering\includegraphics{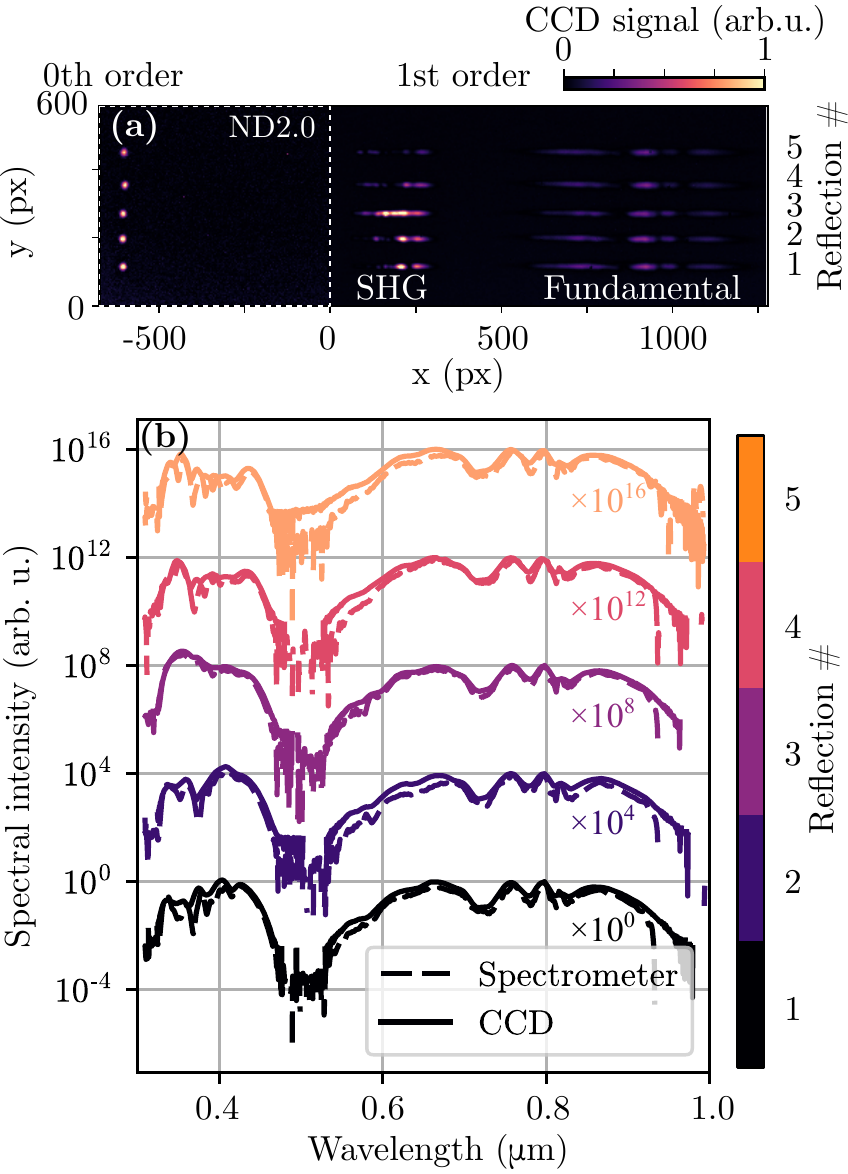}
\caption{(a) Raw CCD image, taken with 1 ms exposure. (b) The comparison of spectra extracted from the CCD image (solid lines) and measured independently with a spectrometer (dashed lines) was used for the wavelength and intensity calibration.\label{fig:camera_read}}
\end{figure}

\section{Experimental setup}

To demonstrate the viability of our method, we assembled the experimental setup shown in \figrefs{fig:setup}. The output of a Ti:Sa laser was spectrally broadened in an Ar-filled hollow-core fiber and compressed to a duration of 6~fs FWHM with a chirped-mirror compressor. For the dispersive plates, we used $M=5$ identical 0.5~mm-thick fused silica windows, which had a wedge of $10^\prime$ to avoid the hindrance by the reflections from their back surfaces. Alternating the direction of the wedge allowed us to minimize the angular chirp induced in the plates. We used a $10~\upmu$m type-I barium borate (BBO) crystal supporting phase-matching across the whole laser bandwidth for SHG, and a 300 grooves/mm UV transmission grating, for spectrally dispersing the beam onto a 8-bit monochrome CCD (Thorlabs) placed 30 mm behind the grating.

An example of an image, acquired by the CCD, is shown in \figref{fig:camera_read}{a}. It consists of two parts, taken with two different laser shots and sewn together at $\mathrm{x}=0$~px. The left part, $\mathrm{x}<0$~px, represents the zeroth order of the diffraction grating, corresponding to the non-diffracted portion of the beams focused on a CCD. The lowest and the highest spot are the reflections from the front surface of the first and last plate, respectively. The intensity decrease from the first to the last beam is consistent with the gradual beam energy loss due to reflections. The zero-order image was taken with an additional neutral density filter placed in front of the CCD, absorbing 99\% of the laser intensity to prevent saturation of the CCD. The camera was then translated to the right ($\mathrm{x}>0$~px), to capture the first diffraction order. Each beam produced two separate traces in this image. The one spanning the region $600$~px~$\lesssim\mathrm{x}\lesssim1200$~px changes only slightly from beam to beam and corresponds to the spectrum of the fundamental beam. A much larger variation can be observed in the second harmonic spectra spanning the region $0$~px~$\lesssim\mathrm{x}\lesssim400$~px. This variation comes from the additional spectral phase that each beam picks up relative to the previous one. In the future, we will use a larger CCD sensor to capture both orders of the grating in the same laser shot. The zeroth order can then be used to account for the beam misalignment on the fly.

\section{Results and discussion}

We recorded the spectrum of each beam separately using a calibrated spectrometer. These spectra (see dashed lines in \figref{fig:camera_read}{b})  were used to calibrate the wavelength and intensity of the spectra acquired on the CCD. We first integrated each trace over the y dimension. We then used the diffraction grating dispersion formula $x(\lambda)=\frac{L}{h}\tan{\left(\arcsin{\frac{\lambda}{d_\mathrm{gr}}}\right)}-x_0$ and substituted it into the resulting signal profile $I_m^\mathrm{CCD}(x)$. Here, $\lambda$ is the wavelength, $L$ is the distance between the grating and the CCD, $h$ is the CCD pixel period, $d_\mathrm{gr}$ is the grating period, and $x_0$ is the pixel position of the zeroth-order beam. We fit a single parameter $L$ to match the positions of characteristic features of $I_m^\mathrm{CCD}(x(\lambda))$ and the spectra acquired with the spectrometer. We also multiplied $I_m^\mathrm{CCD}(x(\lambda))$ by a smooth sensitivity function to account for different spectral sensitivity of the two methods. The resulting spectra are shown in \figref{fig:camera_read}{b}. Comparing the two we estimated the spectral resolution at 5~nm. 

\begin{figure}[ht!]
\centering\includegraphics{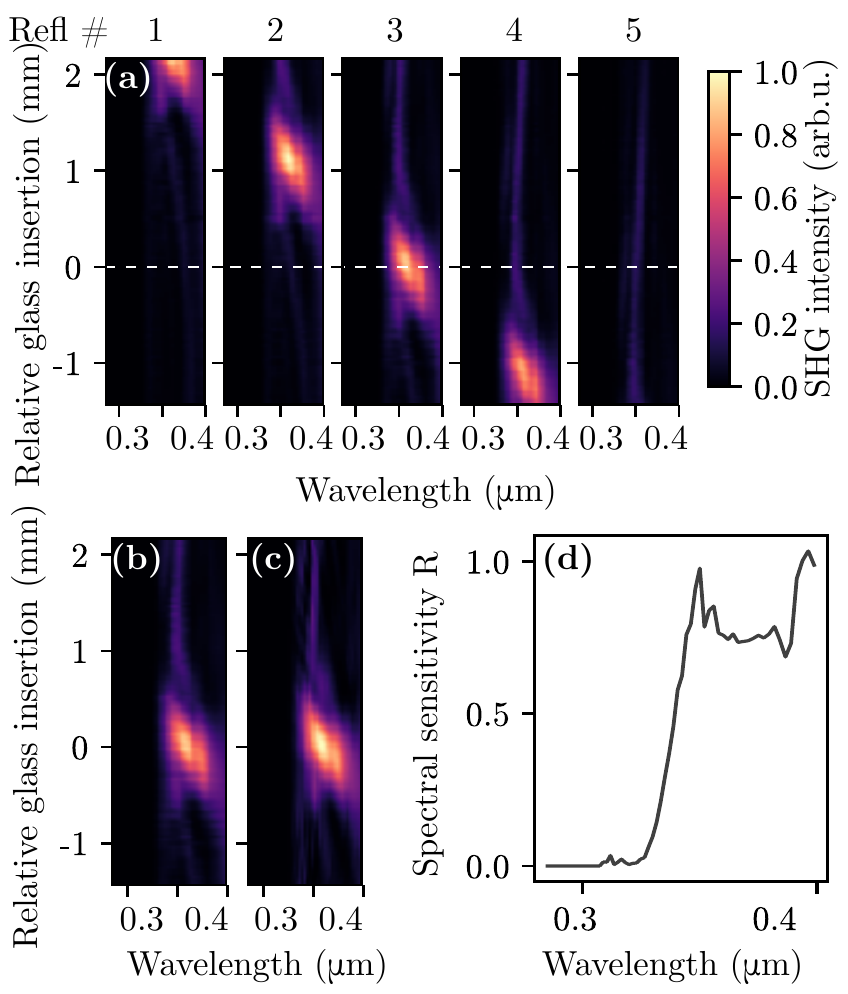}
\caption{(a) Conventional multi-shot dispersion scan spectrogram, recorded for the reflection from all five plates simultaneously. (b)-(c) Comparison of the measured (b) and reconstructed (c) spectrograms. (d) Spectral sensitivity of the detection system retrieved from the full dispersion scan.\label{fig:calib}}
\end{figure}

To assess the accuracy of our method, we compared it to the conventional dispersion scan. We introduced a pair of fused silica wedges into the beam path, and recorded the SHG spectra of the five beams as a function of the beam propagation distance through the glass. The latter is measured relatively to the glass insertion required for the optimal compression of the Reflection \#3 in the resulting spectrograms shown in \figref{fig:calib}{a}. Since the $m$-th beam passes an additional 1~mm of fused silica compared to the $(m-1)$-th, each spectrogram is identical to its neighbors up to a shift by 1~mm along the glass thickness coordinate. We applied the DE reconstruction to the spectrogram of beam 3. The measured and reconstructed traces are shown in \figref{fig:calib}{b, c}, respectively, and demonstrate good agreement, with $G=0.02$.

Single-shot reconstruction involved application of the DE algorithm to the horizontal cross-section of \figref{fig:calib}{a} at 0 relative glass insertion, along the dashed white line. We found that due to the low sensitivity $R(\lambda)$ of the CCD sensor below $\lambda\lesssim350$~nm, simply applying the original procedure\cite{Miranda2012} results in a large error. While the original dispersion scan method utilizes large data oversampling to find $R(\lambda)$, minimizing the error\cite{Miranda2012}, we found this technique to be inefficient for small $M$ and strongly varying $R(\lambda)$. Instead, we extracted $R(\lambda)$ from the conventional dispersion scan reconstruction result,
$$R(\lambda)=\frac{\sum\limits_m\left\{I_m^\mathrm{SHG, meas}(\lambda)I_m^\mathrm{SHG, recons}(\lambda)\right\}}{\sum\limits_mI_m^\mathrm{SHG, recons}(\lambda)^2},$$
shown in \figref{fig:calib}{d}, and then used the DE algorithm adapted to a fixed $R(\lambda)$ on a single-shot spectrogram.

\begin{figure}[ht!]
\centering\includegraphics{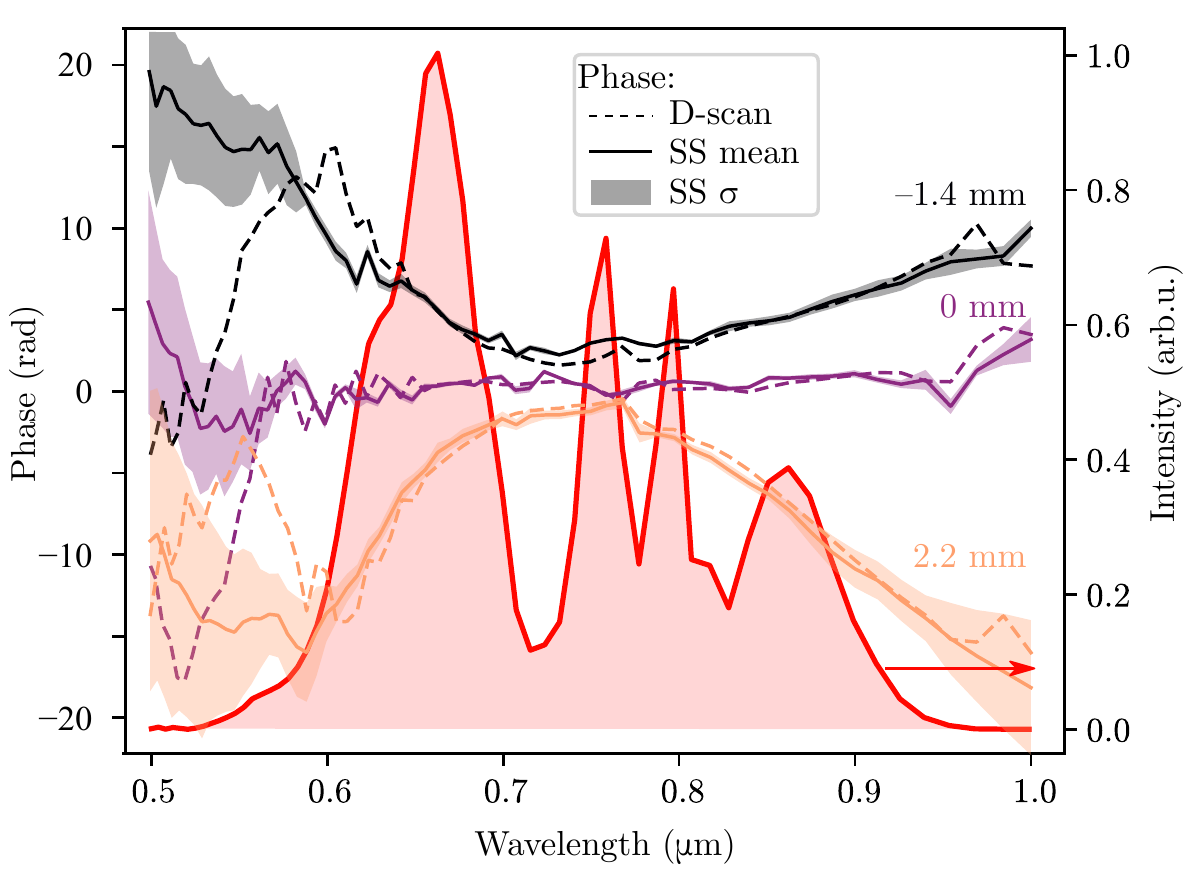}
\caption{Comparison of the pulse phase, retrieved in a multi-shot dispersion scan (dashed lines) and our single-shot method (solid lines and bands of the corresponding colors). The lines and the bands show the average and standard deviation of the phase acquired in 50 consecutive single-shot retrievals. Black, magenta and orange lines correspond to the relative glass insertion of -1.4, 0 and 2.2~mm, respectively. The red line is the spectral intensity of the pulse.\label{fig:reconstr}}
\end{figure}

The results are summarized in \figrefs{fig:reconstr}. We measured the spectral phases of three different pulses with the same spectral intensity (in red): the pulse compressed after the reflection from plate \#3 (magenta), and the two pulses obtained by positively (orange) or negatively (black) chirping it. The chirp was obtained by adding 2.2~mm or removing 1.4~mm of fused silica plates to the beam path, respectively. First, we characterized the spectral phase of each pulse with the conventional dispersion scan (dashed lines). We then carried out single-shot reconstruction of the same data 50 times in a row, to estimate the error in the acquired values. The resulting ranges are shown as shaded regions, the solid lines representing the average phase over 50 reconstructions. We observed a good agreement between our method and the conventional dispersion scan, as well as a negligible variation in different runs of the single-shot algorithm. 

\section{Conclusion}
In conclusion, we have demonstrated a compact and inexpensive method for real-time single-shot reconstruction of optical pulses. Based on a dispersion-scan technique, it extends single-shot reconstruction to an arbitrary dispersion profile, paving the way to universal devices, suitable for the measurement of pulses in a broad range of wavelength and bandwidths. Compared to previously reported single-shot reconstruction techniques, our technique is compatible with a high acquisition rate, and is robust to beam profile inhomogeneity and misalignment. We  implemented a fast GPU-based parallel algorithm, to achieve a real time reconstruction rate of a few Hz. In the future, the technique can be made even more compact and robust, by stacking the plates into a single monolithic device. Moreover, with most of the beam intensity transmitted through the plates available, we could combine it with a single-shot CEP-tagging method, to facilitate the complete characterization of the electric field. With the outlined advantages we believe that our method may become a standard for pulse characterization in the laser laboratories around the world.

\section*{Funding}
Air Force Office of Scientific Research (FA9550-16-1-0109); 
Defense Threat Reduction Agency (HDTRA1-19-1-0026); Canada Research Chairs; Natural Sciences and Engineering Research Council of Canada; Joint Center for Extreme Photonics; Deutsche Forschungsgemeinschaft (KL-1439/11-1, LMUexcellent); Max-Planck-Gesellschaft; European Research Council (FETopen PetaCOM).

\section*{Acknowledgments}
We thank David Crane and Ryan Kroeker for their technical support and are grateful for fruitful discussions with Hartmut Schr\"oder. We thank Gregory Naukhatsky for the provided GPU.

\section*{Disclosures}
The authors declare that there are no conflicts of interest related to this article.








\end{document}